\documentstyle[dina4,12pt,epsfig,here]{article}
\begin{document}

\begin{center}
{\Large {\bf Strange Mesons as a Probe for Dense Nuclear Matter}}

\vspace{1.cm}
Peter Senger for the KaoS Collaboration$^*$

Gesellschaft f\"ur Schwerionenforschung mbH\\ 
Planckstr.1, D-64291 Darmstadt, Germany 

\vspace{1.cm}
{\large {\bf Abstract}}
\end{center}
{\sl
The production and propagation of kaons and antikaons
has been studied in symmetric nucleus-nucleus
collisions in the  SIS energy range.
The ratio of the excitation functions of K$^+$ production
in Au+Au and C+C collisions increases with decreasing beam energy. This
effect was predicted for a soft nuclear equation-of-state.
In noncentral Au+Au collisions, the K$^+$ mesons are preferentially emitted
perpendicular to the reaction plane.
The K$^-$/K$^+$ ratio from A+A collisions at beam energies which are 
equivalent with respect to the threshold is found to be
about two orders of magnitude larger than the corresponding ratio from
proton-proton collisions. 
Both effects are considered to be experimental signatures for
a modification of kaon properties in the dense nuclear medium.
}

\section{Introduction}

Relativistic heavy ion collisions provide
the unique possibility to study nuclear matter at high densities.
At bombarding energies of 1 - 2 AGeV, baryonic densities of 2 - 3 times
saturation density are reached over a timespan  of about 15 fm/c.   
The abundances and phase-space distributions of produced particles
contain information on the properties of the fireball
and thus can be linked to the compressibility of nuclear matter
\cite{aich_ko,maruy,li_ko_fa}. In particular 
K$^+$ mesons are well suited to probe the hot and dense stage of
a nucleus-nucleus collision because of their long mean free path in 
nuclear matter.

The properties of kaons and antikaons are expected to change significantly
inside the dense and hot nuclear medium \cite{kaplan,klimt,brown91}.
The effective mass of K$^+$ mesons is predicted to increase
weakly with increasing nuclear density whereas the effective
mass of the antikaon strongly decreases. 
The latter effect may have consequences 
for the maximum size of neutron stars and the formation of low mass 
black holes \cite{brobet}.
Experiments on kaon and antikaon
production in nucleus-nucleus collisions allow to study both the
nuclear equation-of-state and the properties of hadrons in the dense
nuclear medium.
In this  article we concentrate on data which have been measured recently
with the Kaon Spectrometer \cite{senger}
at the heavy ion synchrotron (SIS) at GSI
in Darmstadt.

\section{Subthreshold kaon production in nucleus-nucleus collisions}
                                                                       
\begin{figure} [hpt]
\vspace{-.5cm}
\begin{minipage}[c]{0.5\linewidth}
    \centering
    \mbox{\epsfig{file=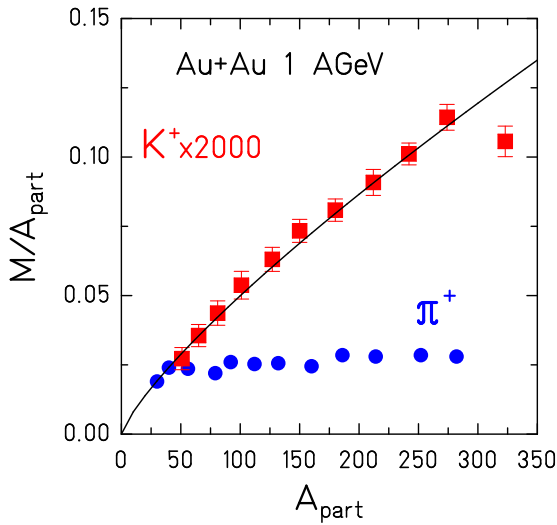,width=10.cm}}
  \end{minipage}\hfill
  \parbox[c]{0.47\linewidth}{
\caption[]{{\sl 
K$^+$ and $\pi^+$ multiplicity per participating nucleon
M/A$_{part}$ as a function of A$_{part}$ for Au+Au collisions at 1 AGeV
\protect\cite{wagner1,mang}. The data are taken
at $\Theta_{lab}$=44$^0$ and extrapolated to the full solid angle
assuming an isotropic angular distribution in the center-of-mass system.
The line corresponds to a parameterization according to
M$_{K^+}\propto$A$_{part}^{1.8}$.
}}
\label{kp_mult_au}
}
\end{figure}

The kaon production threshold in free nucleon-nucleon collisions
is  E$_{beam}$ = 1.58 GeV (for NN$\rightarrow\Lambda$K$^+$N).
In nucleus-nucleus collisions, however,  K$^+$  production is possible 
at beam energies far below this value. 
Fig.~\ref{kp_mult_au} shows the K$^+$ and $\pi^+$ multiplicity per 
participating nucleon M/A$_{part}$ for
Au+Au collisions at 1 AGeV as function of A$_{part}$.
The K$^+$ multiplicity scales with A$_{part}$ according to
M$_{K^+}\propto$A$_{part}^{1.8\pm0.15}$ whereas the $\pi^+$ multiplicity 
scales approximately linearly with A$_{part}$.
The increase of M$_{K^+}$/A$_{part}$ with increasing size of the
collision system is an experimental signature for
kaon production via collective effects. 
Within the framework of transport models calculations, 
''subthreshold''  kaon production
predominantly proceeds via sequential  processes involving intermediate
pions or $\Delta$ resonances: in a first step a $\Delta$ (or a pion)
is produced which collides subsequently with a baryon and produces
a kaon: $\Delta$N$\rightarrow$KYN or $\pi$N$\rightarrow$KY
with Y=$\Lambda,\Sigma$  \cite{aich_ko,fuchs97,cass97}.
In these processes, the $\Delta$ resonance or the pion 
serves as an energy reservoir which
significantly lowers the K$^+$ production threshold.

\vspace{-.5cm}
\section{Probing the nuclear equation of state}
The equation-of-state of nuclear matter (EOS) plays an important role 
for the stability of 
neutron stars or the dynamics of a supernova explosion. The observation,
that the masses of neutron stars $(3\rho_0 < \rho < 10\rho_0)$
have values around 1.5 solar masses
excludes a very stiff EOS. The prompt explosion
mechanism of a supernova ($\rho\leq 4\rho_0$) is only possible for 
a very soft EOS.
However, if the star has angular momentum or if the shock wave is 
revived by neutrino heating, the explosion may also happen for a stiffer EOS
\cite{hillebrandt}.   
Therefore, information on the EOS can hardly be extracted from 
astrophysical observations but rather is needed as an input for stellar models. 

Properties of nuclear matter can be studied in laboratory experiments. 
The analysis of data on the giant monopole resonance (''breathing mode'')
in heavy nuclei finds compressibilities of $\kappa$ = 210$\pm$30 MeV
\cite{blaizot}. However, in this case the nuclear density varies only
by less than 1\% around $\rho_0$. From  data on the refractive
elastic scattering of two oxigen nuclei at bombarding energies of
9 - 30 AMeV values of $\kappa$ = 170-270 have been extracted \cite{khoa}.
In relativistic heavy ion collisions at SIS energies, nuclear matter is
compressed up to baryonic densities of $\rho\approx 3\rho_0$.
The decompressional collective motion of nucleons and light fragments
is related to the stiffness of the EOS. 
However, the analysis  of data on the directed
flow of nucleons into the reaction plane 
has not yet resulted in an unambiguous 
determination of the nuclear compressibility because of the influence
of the momentum dependence of the nucleon-nucleon interaction on this
observable \cite{aichelin}. Recent theoretical studies predict a
robust sensitivity of the ''elliptic'' flow on the EOS \cite{daniel98}.

\begin{figure} [H]
\vspace{-0.5cm}
\begin{minipage}[c]{0.5\linewidth}
    \centering
    \mbox{\epsfig{file=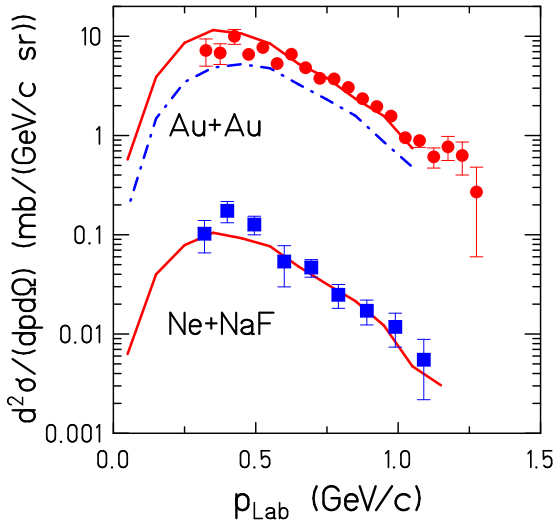,width=10.cm}}
  \end{minipage}\hfill
  \parbox[c]{0.47\linewidth}{
\caption{{\sl
Double-differential K$^+$ production cross section
measured in A+A collisions at 1.0 AGeV  ($\Theta_{lab}$=44$^0$) as
function of laboratory momentum (circles: Au+Au \protect\cite{mang},
squares: Ne+NaF \protect\cite{ahner}).
The lines represent results of RBUU calculations for a soft 
($\kappa$= 200 MeV, solid line) and
a hard ($\kappa$= 380 MeV, dashed-dotted line) 
equation of state \protect\cite{fang_ko}.
For the Ne+NaF system no difference between soft and hard EOS is visible.
}}
\label{kp_ne_au}
}
\end{figure}                    

Data on particle production in nucleus-nucleus collisions have also been 
analyzed with respect to nuclear matter properties. The first attempt 
to determine the nuclear compressibility  via meson observables  
was  made using   the total pion multiplicity  
as a thermometer \cite{harris1}.      
Subthreshold K$^+$ production in relativistic nucleus-nucleus collisions
has been considered to be a promissing
probe to study the properties of nuclear matter
at high densities.
The sensitivity of kaon production on matter properties is based on (i)
the collective production processes via multiple interactions which are
strongly enhanced in the dense phase of the collision and on (ii)
the long mean free path of K$^+$ mesons which may serve as nearly
undisturbed messengers.
Transport model calculations  predict that the K$^+$ yield obtained 
in Au+Au collisions at 1 AGeV is about 2 times higher for a soft EOS than for a 
stiff EOS \cite{li_ko}. This is demonstrated in                               
Fig.~\ref{kp_ne_au} which shows experimental results for
two different system sizes: Ne+NaF  and  Au+Au  collisions
at 1.0 AGeV  \cite{mang,ahner}. The data are compared to relativistic
transport calculations (RBUU) with a soft (solid lines) and stiff EOS
(dashed line) \cite{li_ko,fang_ko}.
For the light system Ne+NaF no difference is visible between a stiff 
and a soft EOS. However, the calculations use  a parameterization of the
NN$\rightarrow$KYN process \cite{ran_ko} which was found to give a too large
cross section
near threshold. Moreover, the process  $\pi$N$\rightarrow$KY is neglected.
Therefore, the absolute agreement of calculations and data as shown in
fig.~\ref{kp_ne_au} should not be overinterpreted.
On the other hand,
the theoretical uncertainties affect both the light and the heavy collision
system in a similar way. Hence the relative agreement of the
model calculation with both the Ne+NaF and the Au+Au data
favors  a soft EOS.

\begin{figure}[hpt]
\vspace{0.cm}
\centerline{
\hspace{0.cm}\mbox{\epsfig{file=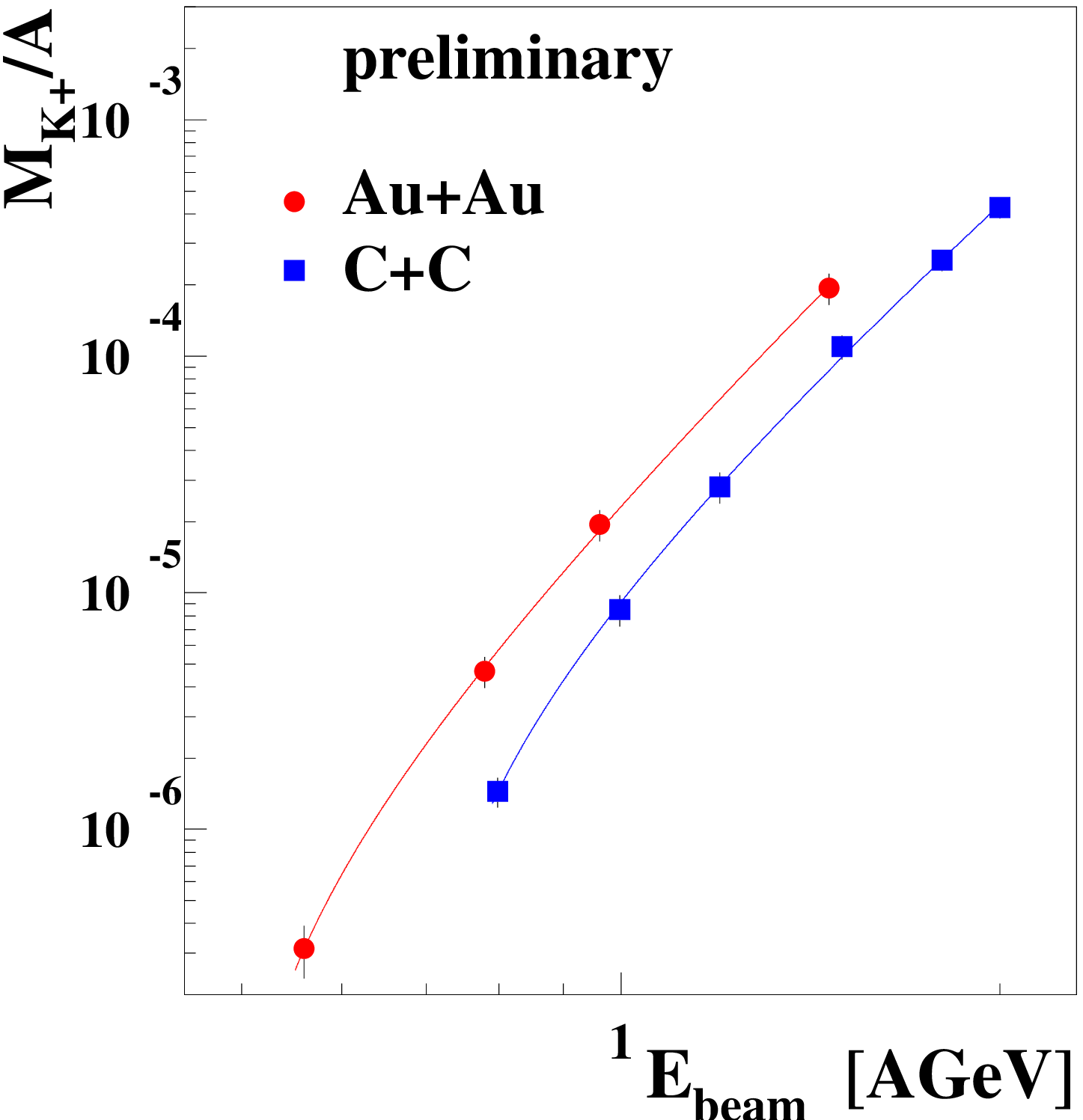,width=8.cm}}
\hspace{0.cm}\mbox{\epsfig{file=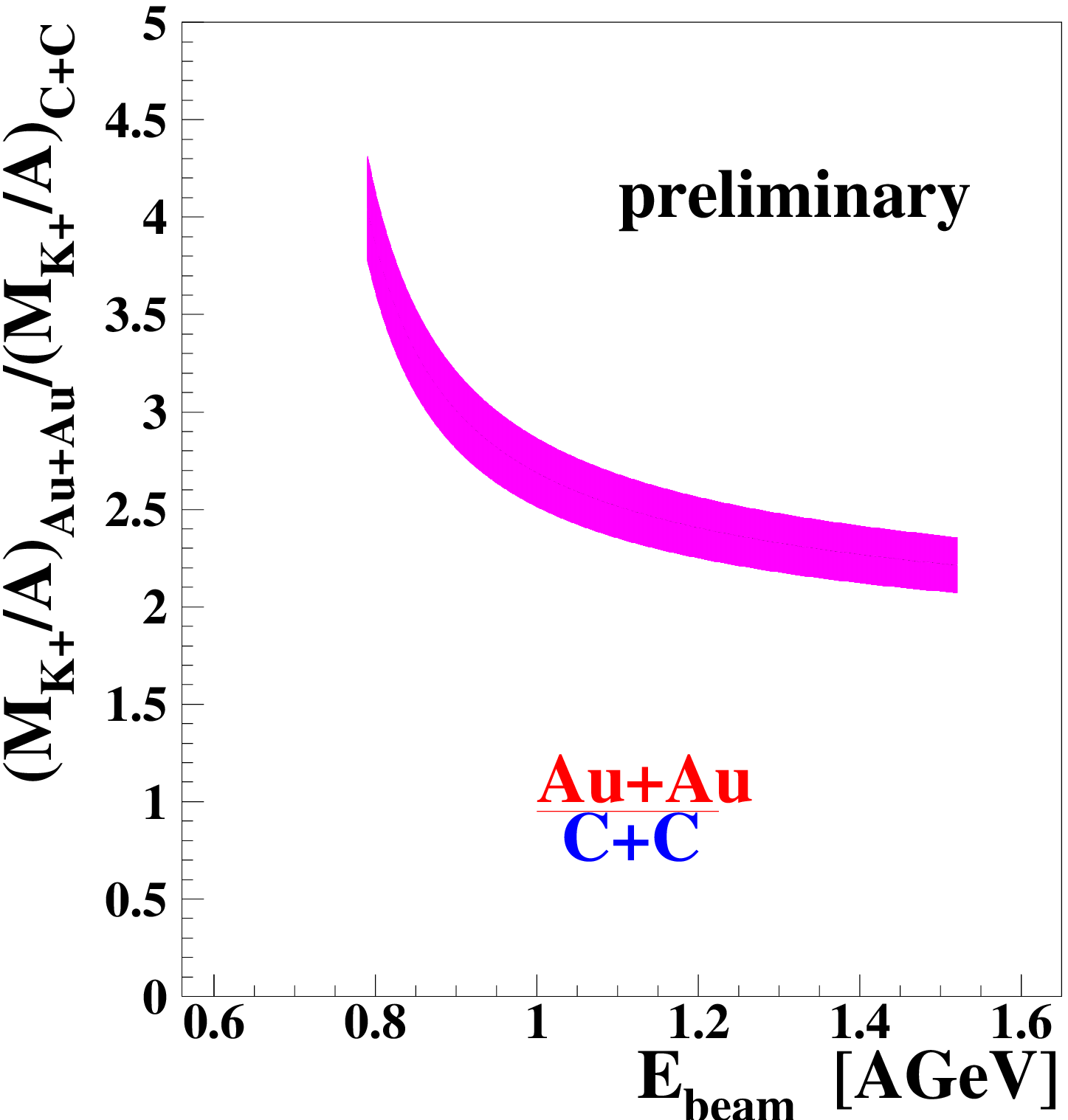,width=8.cm}}
}
\caption[]{{\sl 
Left: excitation function of the K$^+$ multiplicity per nucleon
for Au+Au und C+C reactions.
Right: ratio of the K$^+$ multiplicity per nucleon
for Au+Au und C+C reactions as function of the beam energy
(preliminary results) \protect\cite{sturm} .
}}
\label{k_exci_au_cc}
\end{figure}

Fig.~\ref{kp_ne_au} demonstrates that the K$^+$ yield is sensitive to 
the nuclear compressibility for heavy collision systems at subthreshold
beam energies. The sensitivity of the  K$^+$ yield to the EOS vanishes  
for light systems and at beam energies well above 
the kaon production threshold \cite{hartnack}. 
Therefore, the excitation functions for K$^+$ production in Au+Au collisions 
should be different from the one in C+C collisions: the cross section ratio 
$\sigma^{K^+}_{Au+Au}(E_{beam})$/$\sigma^{K^+}_{C+C}(E_{beam})$
is expected to increase with decreasing beam energy $E_{beam}$ for a soft
EOS.  The measured K$^+$ excitation functions are presented in 
the left part of fig.~\ref{k_exci_au_cc} for Au+Au and C+C collisions 
\cite{sturm}.
The total cross sections have been calculated from the differential cross
sections for K$^+$ production which are integrated over momentum and 
extrapolated to the full solid angle assuming a nonisotropic  polar angle 
distribution. This distribution was determined from 
measurements of K$^+$ mesons at different laboratory angles and 
was found to be slightly  forward-backward peaked  
\cite{sturm,laue}. The right part of fig.~\ref{k_exci_au_cc}  
presents the ratio of the excitation functions as given by the lines in the 
left part of the figure. The width of the band correponds to the uncertainty
of the data points. It can clearly be seen that the ratio increases with
decreasing beam energy. However, this cannot be considered yet as a proof 
for a soft EOS. It remains to be studied to what extend the increase of the 
ratio towards lower beam energies is caused by a reduced kaon yield in 
C+C collisions due to the small number of participating nucleons. 
A detailed analysis of these data using state-of-the-art 
transport models should clarify
this question and put 
further contraints on the compressibility of nuclear matter.

\section{In-medium modifications of kaons and  antikaons}
The formation of a nuclear fireball in nucleus-nucleus
collisions provides the possibility to study
the properties of hadrons under extreme conditions.
It has turned out, that the produced K-mesons  are promissing candidates for
the experimental study of in-medium modifications.
The properties of kaons and antikaons in dense nuclear matter
have been investigated using chiral perturbation theory
\cite{kaplan,brown91}, chiral dynamics \cite{klimt},
relativistic mean field models \cite{schaff96} and a self-consistent
coupled-channel
approach \cite{lutz}.
The calculations find an attractive  kaon-nucleon (scalar) potential
which is related to explicit
chiral symmetry breaking due to the large strange quark mass.
The kaon-nucleon vector potential is repulsive for kaons
but attractive for antikaons. Hence
the total KN interaction in the medium  is
weakly repulsive for kaons but strongly
attractive for antikaons.
These in-medium KN potentials influence the propagation of kaons and antikaons
in nuclear matter. As a consequence, their azimuthal emission pattern
is expected to be modified according to the density profile of the nuclear
medium: K$^+$ mesons will be repelled from the regions of increased baryonic
density whereas K$^-$ mesons will be attracted \cite{li_ko_br}
\begin{figure}[H]
\vspace{-0.cm}
    \centerline{
\hspace{-.5cm}\mbox{\epsfig{file=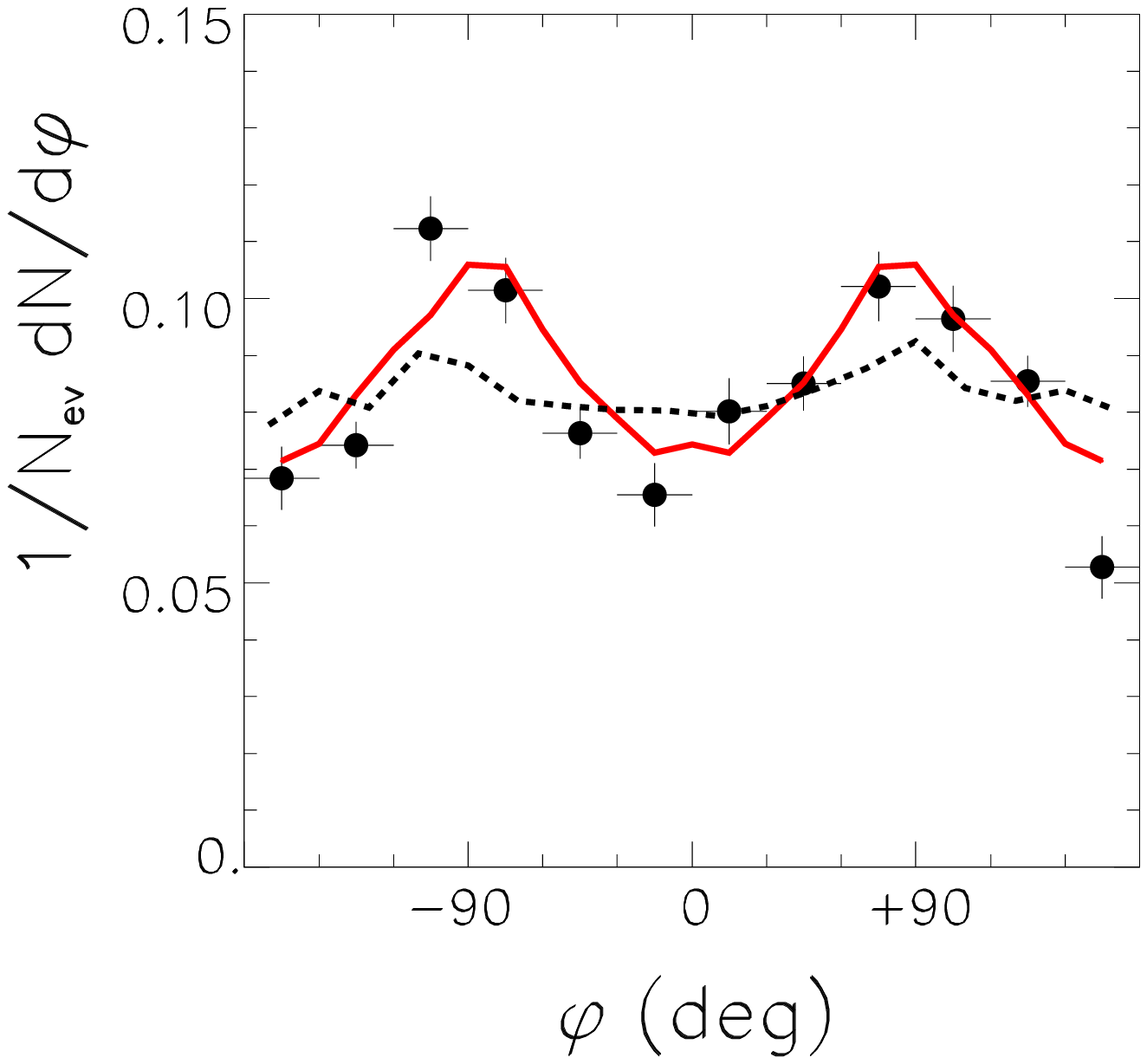,width=6.cm}}
\hspace{2.cm}\mbox{\epsfig{file=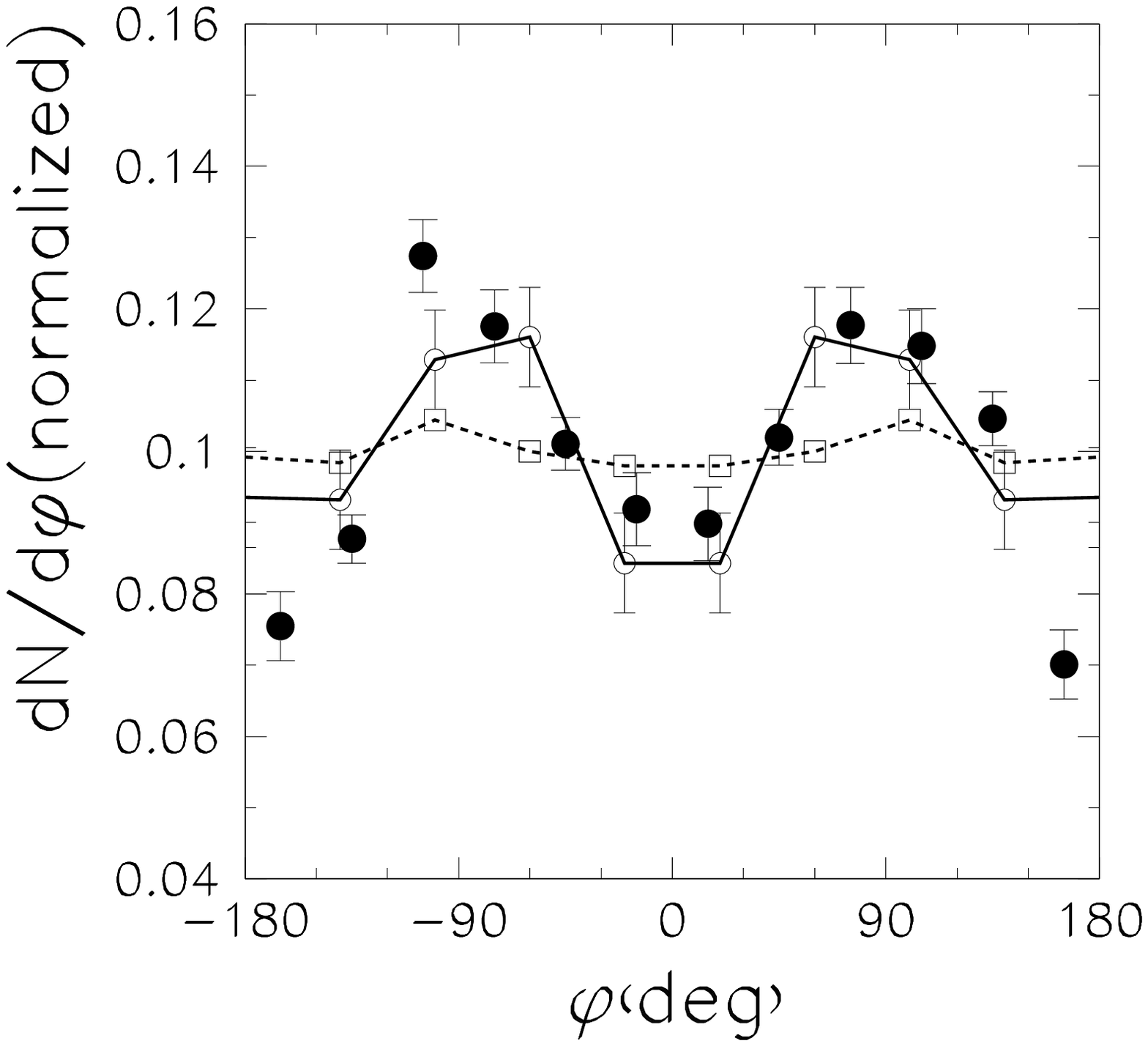,width=6.5cm}}
} 
\vspace{-0.cm}
\caption[]{{\sl
K$^+$ azimuthal distribution for semi-central Au+Au collisions at 1 AGeV
(full dots). The data are analyzed at $0.4< y/y_{proj}< 0.6$ (left) 
and $0.2< y/y_{proj}< 0.8$ (right) \protect\cite{shin}.
The lines represent results of transport calculations using a RBUU model
(left \protect\cite{li_ko_br}) and a 
QMD model (right \protect\cite{wang}). Both models take into account 
rescattering, the QMD version also considers Coulomb effects.
Solid lines: with in-medium KN potential.
Dashed lines: without  in-medium KN potential.
}}
\label{shin4}
\end{figure}

Fig.~\ref{shin4} presents the K$^+$ azimuthal angular distribution measured in
Au+Au collisions at 1 AGeV \cite{shin}. The kaons were accepted within a range
of transverse momenta of 0.2 GeV/c $\leq$ p$_t$$\leq$ 0.8 GeV/c
for two ranges of normalized rapidities  
0.4 $\leq$ y/y$_{proj}$ $\leq$ 0.6 (left) and 
0.2 $\leq$ y/y$_{proj}$ $\leq$ 0.8 (right). 
The data are corrected for the uncertainty of the 
determination of the reaction plane by a Monte Carlo simulation.
The K$^+$ emission pattern clearly is peaked at $\phi$=$\pm$90$^0$
which is perpendicular to the reaction plane. Such a behaviour is known from
pions \cite{brill} which interact with the spectator fragments. 
The K$^+$ mesons, however,
have a long mean free path of about 5 fm and therefore are less
sensitive to rescattering. This  is demonstrated in
fig.~\ref{shin4} by the dotted lines which represent the results of a
transport calculation taking into account K$^+$ rescattering only 
(\cite{li_ko_br}, left) and additional Coulomb effects (\cite{wang}, right). 
However, if a repulsive in-medium KN potential is assumed, 
the calculations reproduce  the
pronounced anisotropy of the data (solid lines in fig.~\ref{shin4}).

\begin{figure}[H]
\vspace{-.5cm}
  \begin{minipage}[c]{0.6\linewidth}
    \centering
\mbox{\epsfig{file=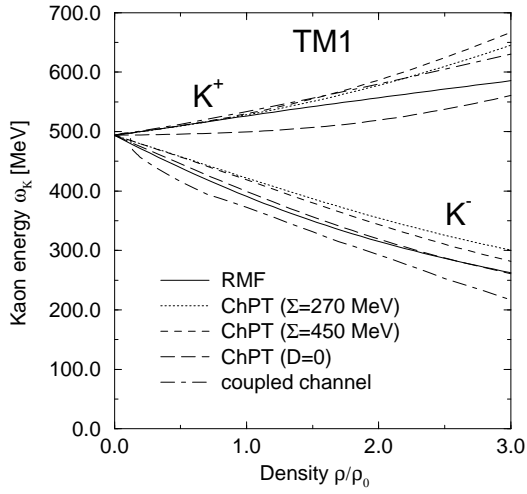,width=7.cm}}
  \end{minipage}\hfill
  \parbox[c]{0.35\linewidth}{
\caption[]{{\sl
Effective in-medium mass of kaons and antikaons
as function of nuclear density according to a different
calculations. Taken from \protect\cite{schaff96}.
}}
\label{kamed}
}
\end{figure}

Another manifestation of the in-medium KN potentials is a modification
of the K$^+$ and K$^-$ effective mass in nuclear matter.
Fig.~\ref{kamed} shows the effective mass of kaons and antikaons
as function of nuclear density as calculated by various models 
\cite{schaff96}. The different calculations exhibit a common trend:
with increasing nuclear density the K$^+$ effective mass 
increases weakly  whereas the
K$^-$ effective mass decreases considerably.

\begin{figure}[hpt]
\vspace{-1.cm}
 \begin{minipage}[c]{0.6\linewidth}
    \centering
\mbox{\epsfig{file=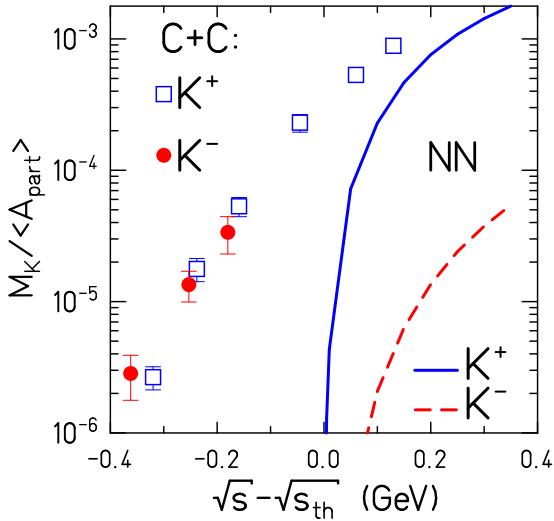,width=10.cm}}
  \end{minipage}\hfill
  \parbox[c]{0.38\linewidth}{
\caption[]{{\sl
Kaon and antikaon multiplicity per participating nucleon
as a function of the Q-value  for C+C
collisions (open squares: K$^+$, full dots: K$^-$) \protect\cite{laue}.
The lines correspond to parameterizations of the 
cross sections for K meson production in nucleon-nucleon collisions
(solid line: K$^+$, dashed line: K$^-$)
\protect\cite{sibirtsev,brat_cass,si_ca_ko}.
}}
\label{cc_kp_km}
}
\end{figure}

The reduction of the 
K$^-$ effective mass in the dense nuclear medium as indicated
in fig.~\ref{kamed} lowers  the K$^-$ production threshold
and thus  enhances the K$^-$ production cross section 
in nucleus-nucleus collisions \cite{li_ko_fa,ko_li}.
The enhancement of the K$^-$ yield should be very pronounced
at beam energies below the kinematical threshold (which is
2.5 GeV for the process NN$\to$K$^-$K$^+$NN) because of the steep excitation
function.
We have found  experimental evidence for an enhanced K$^-$ yield
in Ni+Ni collisions at 1.8 AGeV \cite{barth}.
Recent experiments on antikaon production in C+C collisions 
confirm the previous results. Fig.~\ref{cc_kp_km} shows the excitation  
function of K$^+$ and K$^-$ production in C+C collisions as function
of the Q-value in the NN system \cite{laue}. 
The kaon and antikaon data nearly fall
on the same curve (open and full symbols) in contrast to the 
parameterizations of the nucleon-nucleon data 
(lines). 
The parameterizations are fitted to the available proton-proton data  
and averaged over the isospin channels \cite{sibirtsev,brat_cass,si_ca_ko}.
The multiplicities are calculated from the production cross sections 
via  M$_K$ = $\sigma_K/\sigma_R$ with the total reaction cross section
$\sigma_R$=0.95 b for C+C and $\sigma_R$=47 mb for p+p. The average number of 
participants is assumed to be $<$A$_{part}>$=6 for C+C (according 
to a geometrical model) and $<$A$_{part}>$=2 for p+p.
The large difference in the K$^+$/K$^-$ ratio for C+C and p+p provides 
strong experimental evidence for an enhanced antikaon production in 
nucleus-nucleus collisions.

\begin{figure}[hpt]
\vspace{-0.cm}
\centerline{
\hspace{0.5cm}\mbox{\epsfig{file=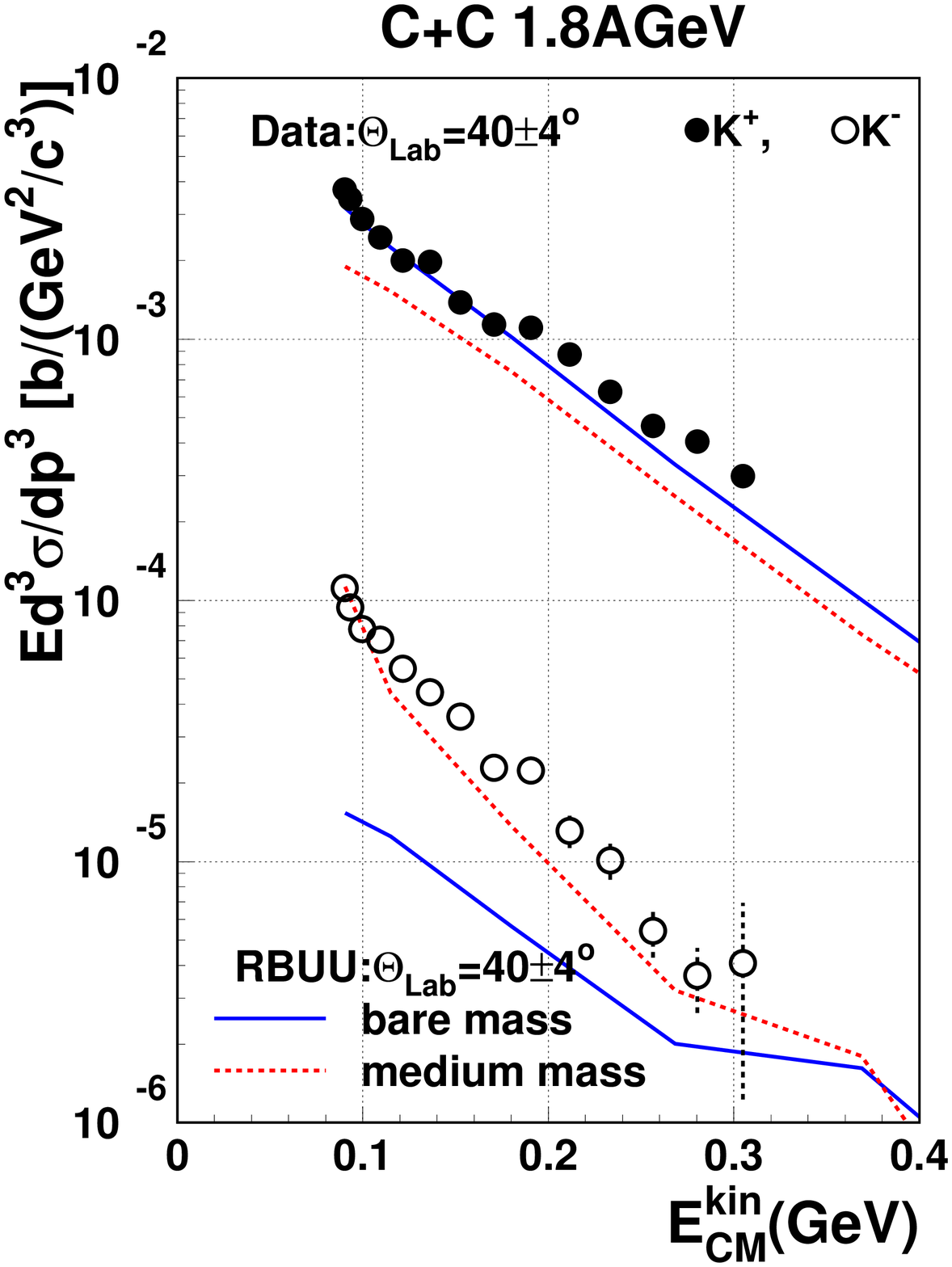,width=7.cm}}
\hspace{-0.5cm}\mbox{\epsfig{file=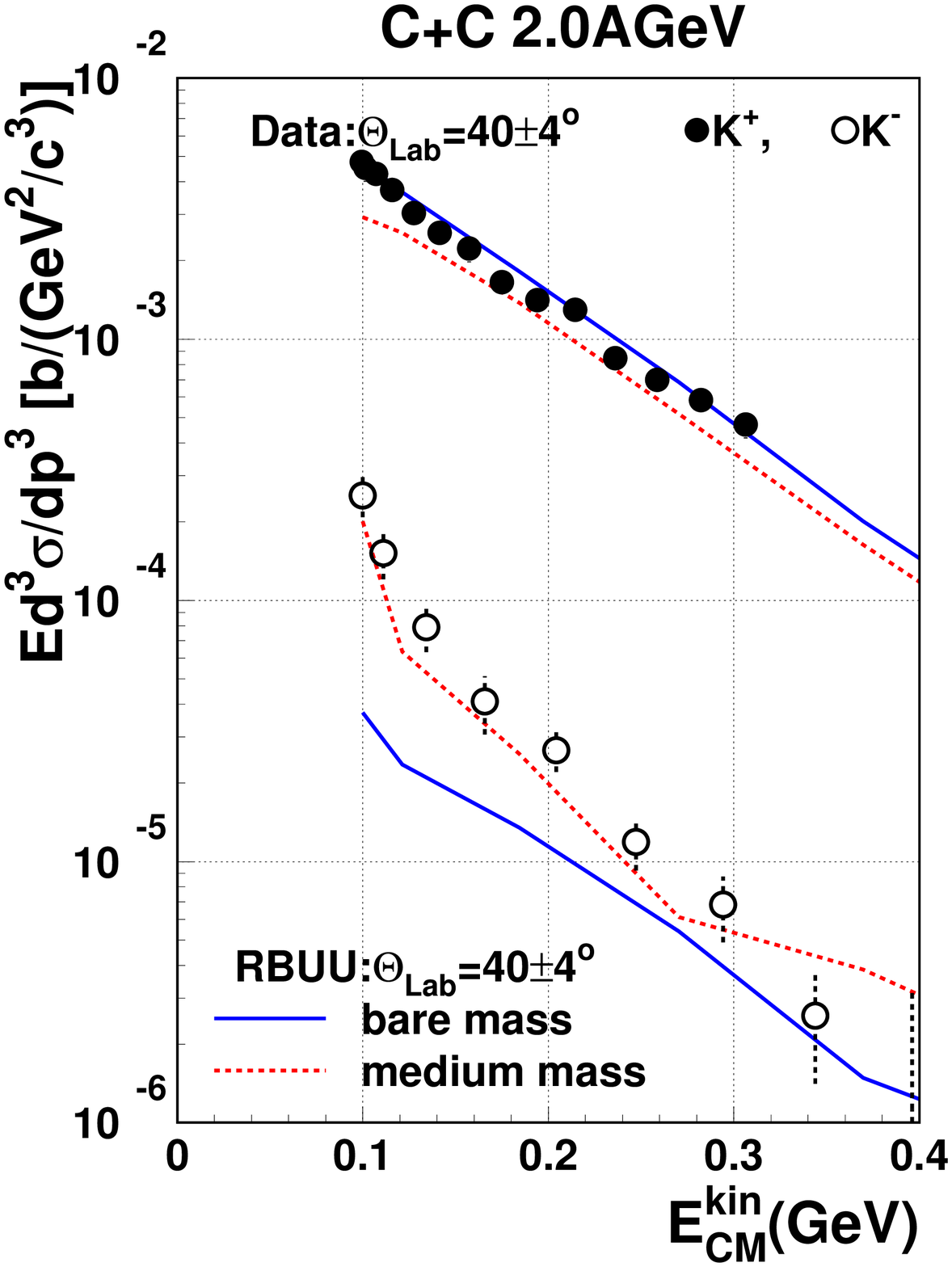,width=7.cm}}
}
\vspace{-0.5cm}
\caption{{\sl Invariant cross sections for kaon and antikaon production
in C+C collisions at 1.8 AGeV (left) and 2.0 AGeV (right) as function 
of the kinetic energy in the center-of-mass system \protect\cite{laue}.
The data were taken near midrapidity. The lines represent predictions 
of RBBU transport calculations assuming bare K meson masses 
(solid lines)  and modified K meson masses (dashed lines)
\protect\cite{brat}.
}}
\label{cc_km_inv}
\end{figure}

Transport models predict a significant enhancement of the K$^-$ yield 
even in light collision systems such as C+C. 
Fig.~\ref{cc_km_inv} shows the K$^+$ und K$^-$ spektra measured in
C+C collisions at beam energies of 1.8 AGeV and 2.0 AGeV 
\cite{laue} in comparison to RBUU results \cite{brat}.
The calculations were performed with different assumptions: 
(i) with  bare masses of the K mesons (solid lines) and (ii) with
modified effective masses according 
to m$^*$ = m$^0$ (1 - $\alpha$ $\rho/\rho_0$) with $\alpha$=0.24 for K$^-$ 
and $\alpha$=-0.06 for K$^+$ (dashed lines). 
Both the calculated and measured kaons were 
taken at $\Theta_{lab}$=44$\pm$4$^0$.   
The K$^+$ data are not very sensitive to the 
variation of the in-medium masses and are compatible with the assumption
of bare masses. The antikaon spectrum, however, shows a very distict dependence
on the K$^-$  effective mass in the medium. The K$^-$ data clearly favor the 
calculation based on the reduced effective mass.

\begin{figure}[hpt]
\vspace{0.cm}
\vspace{-0.cm}
  \begin{minipage}[c]{0.55\linewidth}
    \centering
\mbox{\epsfig{file=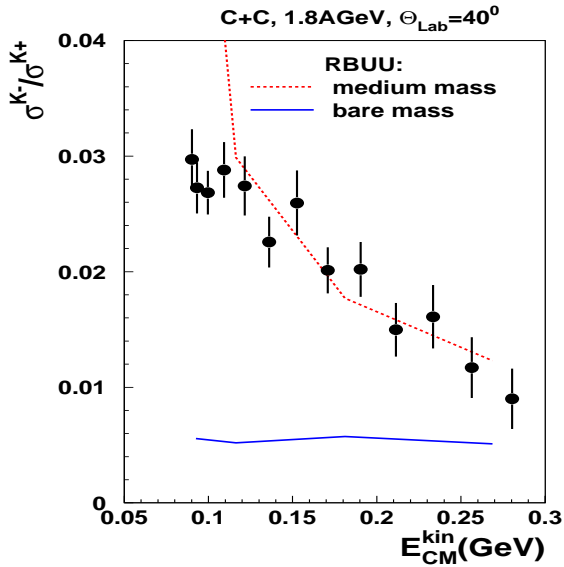,width=8.cm,height=8cm}}
  \end{minipage}\hfill
  \parbox[c]{0.40\linewidth}{
\caption{{\sl K$^-$/K$^+$ ratio as function of the kinetic energy in the 
center-of-mass system  for C+C collisions at 1.8 AGeV \protect\cite{laue}.
The lines represent predictions od RBUU calculations.
Solid lines: without in-medium effects (bare mass),
dotted line:  with in-medium effects (medium mass)
\protect\cite{brat}.
}}
\label{karatcc18_buu}
}
\end{figure}

The systematic uncertainties of the calculations and of the experimental
data are reduced when looking at K$^-$/K$^+$ spectral ratios.
Fig.~\ref{karatcc18_buu} presents  the K$^-$/K$^+$  ratio as function of the
kinetic energy for C+C collisions at 1.8 AGeV in comparison to the RBUU 
results (same as in fig.~\ref{cc_km_inv}).
The K$^-$/K$^+$ ratio steeply
decreases with increasing kinetic energy of the K mesons.
This is not a trivial observation
as K$^-$ mesons with low energies are  stronger affected by absorption than
K$^-$ mesons with  higher energies \cite{dover}.
Relativistic transport calculations predict a constant  K$^-$/K$^+$ ratio
if in-medium mass modifications
of the K mesons are neglected (solid line in fig.~\ref{karatcc18_buu}). 
The model calculations are in 
good agreement with the data if in-medium effects are taken into account
(dotted line). 

\section{Conclusions and outlook }
We have presented data on kaon and antikaon production in nucleus-nucleus
collisions at SIS energies. Emphasis was put on the study of in-medium effects 
using strange mesons as probes.
We propose to use the ratio of the excitation functions for K$^+$ production
in very heavy and light systems as an indicator which is sensitive to 
the compressibility
of nuclear matter at high densities. A systematic analysis of the 
K$^+$ yield as function of system mass and beam energy
within the framework of transport models allows
to reduce the uncertainties due to the reaction mechanism  
(cross sections, momentum dependent interactions)
and  enhances the sensitivity of the kaon yield on nuclear matter properties. 
The comparison of the data to published results of transport codes 
favor a  soft EOS.    

In noncentral Au+Au collisions at 1 AGeV we found   
that K$^+$ mesons are emitted preferentially perpendicular to
the reaction plane. A similar effect was observed for pions  
and attributed to rescattering at the spectator matter. The K$^+$ mesons, 
however, have a larger mean free path in nuclear matter and should be
much less affected by the spectators. Transport calculations have to assume
a repulsive kaon-nucleon potential in the medium in order to reproduce the 
azimuthal anisotropy of kaon emission.  
  
We found experimental evidence for an enhanced production of  antikaons
in nucleus-nucleus collisions at beam energies below the free NN threshold. 
The K$^-$/K$^+$ ratio determined for the same Q-value 
($\sqrt s$-$\sqrt s_{thresh}$) in Ni+Ni or C+C collisions exceeds
the ratio measured in proton-proton collisions by about two orders
of magnitude. Transport models cannot explain the 
yield of antikaons when neglecting in-medium modifications of the K$^-$ meson.
The strangeness exchange reaction Y$\pi\to$K$^-$N (with Y=$\Lambda,\Sigma$)
produces  only about 10\% of the measured K$^-$ yield, another 10\%
is due to the process $\pi$N$\to$K$^+$K$^-$N. Only about 1 \% of the K$^-$
mesons stem from nucleon-nucleon collisions NN$\to$K$^+$K$^-$NN.  
However, if the effective mass of the antikaon  drops with increasing  
baryonic density, the in-medium thresholds for all these
processes are reduced and the total K$^-$ yield is enhanced by a 
factor of about 5 \cite{cass97,brat_cass}. Recent coupled-channel calculations
find that the  in-medium effects are reduced 
with increasing momentum of the antikaon \cite{lutz}.   

The behaviour of strange particles in dense nuclear matter
is expected to play an important  role in astrophysics. 
It is still an open question why the masses of the observed pulsars  
dont exceed a value of  about 1.5 solar masses.
In order to explain this upper limit, stellar evolution models would need 
a high density EOS which is much softer than generally thought.
Glendenning and Weber suggested that pulsars might consist 
of neutrons, protons and hyperons because the EOS of these hybrid stars 
is considerably softer than the one of pure neutron stars  \cite{glen_web}.
Bethe and Brown proposed another mechanism which  softens the EOS of
the compact core of a collapsing  star:
if the effective mass of the K$^-$ meson decreases with increasing nuclear 
density  then the total K$^-$ meson energy  will become  smaller 
than the electrochemical potential ($\mu_e\approx$230 MeV)
above a certain value of the nuclear density. In this situation  the 
K$^-$ mesons may replace
electrons and form a Bose condensate. The condensation of negative
particles enhances the proton to neutron ratio and this effect
softens the EOS. Consequently, a supernova core
with 1.5 - 2 solar masses will collapse
into a black hole rather than  form a neutron star  \cite{brobet}. 
The authors claim  that a prominent example for this scenario is 
Supernova 1987A.  
Recently, Brown and coworkers analyzed  antikaon yields measured 
in Ni+Ni collisions at 1.8 AGeV \cite{barth}. They extracted from the data 
the density dependence of the effective mass of the antikaon and predicted
K$^-$ condensation in neutron stars
above 3 times saturation density $\rho_0$ \cite{li_lee_br}. 

\vspace{1.cm}
$^*$ The KaoS Collaboration

The data presented in this article have been measured and analyzed by
the KaoS Collaboration which actually consists of the following
persons (names of major workers are printed in boldface):

P.Koczo\'n, {\bf F.Laue}, P.Senger, G.Surowka (GSI)

A.F\"orster, H.Oeschler, {\bf C.Sturm},  F.Uhlig (TU Darmstadt)

E.Schwab, {\bf Y.Shin}, H.Str\"obele (Univ. Frankfurt)

I.B\"ottcher, B.Kohlmeyer, M.Menzel, F.P\"uhlhofer (Univ. Marburg)

M.D\c{e}bowski, W.Walu\'s (Jagiell. Univ. Krak\'ow)

F.Dohrmann, E.Grosse, L.Naumann, W.Scheinast (FZ Rossendorf)

A.Wagner (NSCL Michigan State University, USA)


\begin{thebibliography}{80}
\bibitem{aich_ko} J.Aichelin and C.M.Ko, Phys.Rev.Lett. {\bf 55} (1985) 2661
\bibitem{maruy} T.Maruyama et al., Nucl.Phys. {\bf A573} (1994) 653
\bibitem{li_ko_fa} G.Q.Li, C.M.Ko, X.S.Fang Phys.Lett. {\bf B 329} (1994) 149
\bibitem{kaplan} D.B.Kaplan and A.E.Nelson, Phys.Lett. {\bf B 175} (1986) 5
\bibitem{klimt} S.Klimt et al., Phys.Lett. {\bf B 249} (1990) 386
\bibitem{brown91} G.E.Brown et al., Phys.Rev. {\bf C 43 } (1991) 1881
\bibitem{brobet} G.E.Brown and H.A.Bethe, Astro.Jour. {\bf 423} (1994) 659
\bibitem{senger}P.Senger et al., Nucl.Instr.Meth. {\bf A 327} (1993) 393
\bibitem{fuchs97} C.Fuchs et al., Phys. Rev. {\bf C 56} (1997) 216
\bibitem{cass97} W.Cassing et al., Nucl. Phys. {\bf A 614} (1997) 415
\bibitem{wagner1} A.Wagner, PhD Thesis, Techn. Univ. Darmstadt 1996
\bibitem{mang} M.Mang, PhD Thesis, Univ. Frankfurt 1997
\bibitem{hillebrandt} W.Hillebrandt, E.M\"uller und R.M\"onchmeyer,
The Nuclear Equation of State, NATO ASI series. Series B: Physics Vol.216A
edited by W.Greiner and H.St\"ocker, Plenum Press, 1989 p.689
\bibitem{blaizot} J.P.Blaizot, J.F.Berger, J.Decharg\'e, M.Girod,
Nucl.Phys. {\bf A 591} (1995) 435
\bibitem{khoa} D.T.Khoa et al., Phys.Rev.Lett. {\bf 74} (1995) 34
\bibitem{aichelin}  J.Aichelin, Phys.Rep.{\bf 202} (1991) 233
\bibitem{daniel98} P.Danielewicz et al., Phys. Rev. Lett. {\bf 81} (1998) 2438
\bibitem{harris1}J.W.Harris et al., Phys.Lett. {\bf 153 B}  (1985) 377
\bibitem{li_ko} G.Q.Li, C.M.Ko, Phys.Lett. {\bf B 349} (1995) 405
\bibitem{ahner} W.Ahner et al.,  Phys.Lett. {\bf B393} (1997) 31
\bibitem{fang_ko} X.S.Fang, C.M.Ko, G.Q.Li, Y.M.Zheng,
Nucl.Phys. {\bf A 575} (1994) 766
\bibitem{ran_ko}J.Randrup and C.M.Ko, Nucl.Phys. {\bf A 343} (1980) 519
und {\bf A 411} (1983) 537
\bibitem{hartnack} C.Hartnack, J.Jaenicke, J.Aichelin, Nucl.Phys.
{\bf A 580} (1994) 643
\bibitem{sturm} C.Sturm, PhD Thesis in prep., Techn. Univ. Darmstadt
\bibitem{laue} F.Laue et al., Phys. Rev. Lett. in print (nucl-ex/9901005) 
and F. Laue PhD Thesis,  Univ. Frankfurt (1999)
\bibitem{schaff96} J.Schaffner-Bielich, J.Bondorf, I.Mishustin,
Nucl. Phys. A 625 (1997) 325
\bibitem{lutz} M.Lutz, Phys.Lett. {\bf B 426} (1998) 12
\bibitem{li_ko_br}  G.Q.Li, C.M.Ko and G.E.Brown, Phys.Lett. {\bf B 381}
(1996) 71 and G.Q.Li, priv. comm.
\bibitem{shin} Y.Shin et al., Phys. Rev. Lett. {\bf 81} (1998) 1576
\bibitem{wang} Z.S.Wang, C.Fuchs, A.Faessler and T.Gross-Boelting, 
nucl-th/9809043 
\bibitem{brill} D.Brill et al., Z. Phys. {\bf A 355} (1996) 61
and Z. Phys. {\bf A 357} (1997) 207
\bibitem{ko_li} C.M.Ko and G.Q.Li, J.Phys. {\bf G22} (1996) 1673
\bibitem{barth} R.Barth et al., Phys.Rev.Lett. {\bf 78} (1997) 4007
\bibitem{sibirtsev} A.Sibirtsev, Phys.Lett. {\bf B 359}  (1995) 29
\bibitem{brat_cass} E.Bratkovskaya, W.Cassing and U.Mosel, Nucl.Phys.
{\bf A 622} (1997) 593
\bibitem{si_ca_ko} A.Sibirtsev, W.Cassing and C.M.Ko, Z.Phys.
{\bf A 358} (1997) 101
\bibitem{brat} E.Bratkovskaya and W.Cassing, private communication
\bibitem{dover} C.B.Dover and G.E.Walker, Phys.Rep.{\bf 89} (1982) 1
\bibitem{glen_web} N.Glendenning and F.Weber, Phys. Rev. {\bf C 45} (1992) 844 
\bibitem{li_lee_br} G.Q.Li, C.H.Lee and G.E.Brown, Phys. Rev. Lett.
 {\bf 79} (1997) 5214
\end{thebibliography}
\end{document}